# Simulations of dusty plasmas using a special-purpose computer system for gravitational N-body problems

K. Yamamoto[a], Y. Mizuno[a], H. Inuzuka[a], Y. Cao[b], Y. Liu[b], K. Yazawa[b]

[a]*Faculty of Engineering, Shizuoka University, 3-5-1 Johoku, Hamamatsu 432-8561, Japan*
[b]*Hamamatsu Metrix Co.,Ltd, 1-4-10-8 shinmiyakoda, 431-2103, Japan*

Simulations of dusty plasmas were performed using GRAPE-6, a special-purpose computer designed for gravitational N-body problems. The collective behaviour of dust particles, which are injected into the plasma, was studied by means of three-dimensional computer simulations. As an example of a dusty plasma simulation, we have calculated movement of dust particles in plasmas under microgravity conditions. In this simulation, a dust-free region called the “void” is observed in the dust cloud. Another example was to simulate experiments on Coulomb crystals in plasmas. Formation of a Coulomb crystal was observed under typical laboratory conditions. For the simulation of a dusty plasma in microgravity with $3x10^4$ particles, GRAPE-6 can perform the whole operation 1000 times faster than by using a Pentium4-1.6GHz CPU.

## 1. Introduction

In order to achieve high calculation speeds, and to lower the cost of the hardware used for simulations, special-purpose computers have been developed recently. A special-purpose computer is typically a massively parallel processing computer, which has many custom LSI chips specifically designed to calculate a few expressions. By limiting the versatility of the calculation, the hardware complexity can be reduced drastically. Therefore, a special-purpose computer can contain a greater number of processors on one chip compare to a general-purpose computer. GRAPE-6 (G6) is a special-purpose computer for gravitational N-body simulations [1] manufactured by Hamamatsu Metrix Co., Ltd. The G6 can rapidly calculate the gravitational force between particles using its specialized pipeline processors. The peak performance of a G6 unit reaches 985 Gflops. We have used G6 for the calculation of Coulomb interactions, which is the same form of central force as the gravitational force, to accelerate the plasma simulation. We have previously developed three-dimensional electrostatic simulations using G6, and performed simulations of plasma waves, fluctuations, and diffusions. Currently, we are investigating developing simulations of dusty plasma using G6. In addition to the plasma, a dusty plasma contains micro-particles, i.e. dust particles. Since micrometer size particles are usually negatively charged - $10^3$~$10^4$$e$ - in a plasma (here $e$ is the elementary charge), the dusty plasma is easily “strongly coupled”. If specific conditions are fulfilled, the dust particles form an ordered crystalline structure called “Coulomb crystal” or “plasma crystal” [2,3]. Many interesting features of dusty plasmas have been elucidated in laboratory experiments, and various shapes of the Coulomb crystal have been observed, such as the disk, dome, and cone [4]. However, in laboratory experiments, the movement of dust particles and the structure of the Coulomb crystal are restricted due to gravity. In order to study dusty plasmas by suppressing the effect of gravity, experiments under microgravity conditions were carried out [5]. In this experiment, a dust-free region called the “void” [6] was observed in the center of the dust cloud. It was also observed that a part of the dust particles around the void were in a crystalline state. In this paper, we present some simulations of a dusty plasma to simulate these experiments, to examine the basic concept of computing using G6, and the actual performance of G6.

## 2. Calculations using GRAPE-6

A unit of G6 has 32 custom LSI chips, each containing 6 pipeline processors for the calculation of gravitational interactions between particles. Since one pipeline processor works logically as 8 pipelines, G6 has 1536 pipelines per unit. G6

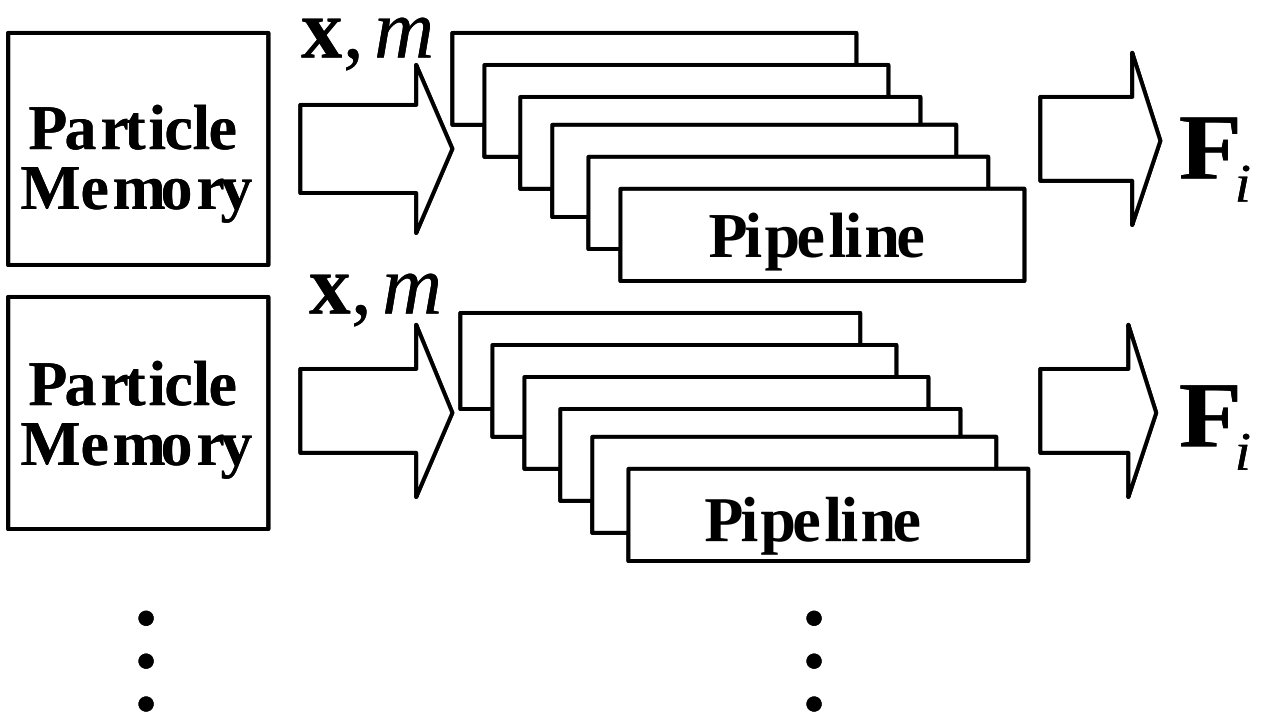


Fig. 1. The architecture of GRAPE-6.

realizes high-speed calculations using these pipelines efficiently in parallel. G6 is connected to a general-purpose host computer (a Pentium 4 personal computer) through a communication interface. G6 calculates only the force due to gravitation between particles, and the host computer performs the other calculations. The user program runs on the host computer, and it can use G6 only by calling library functions. The architecture of G6 is fairly simple from the viewpoint of a user. G6 is comprised of multiple memory units and pipeline processors. Each LSI chip has its own memory unit, and whole particle data is stored distributed to these memory units. Thus, the calculation of the force, which is exerted on a particle, is divided among all LSI chips. Besides, G6 calculates the forces on different particles in parallel using pipeline processors on the LSI chip. The architecture of G6 is shown in Fig.1.

The gravitational force calculated by a pipeline of G6 is given by

$$\boldsymbol{F}_i = G m_i \sum_j m_j \frac{\boldsymbol{r}_{ij}}{(r_{ij}^2 + S^2)^{3/2}} \qquad (1)$$

$$\boldsymbol{r}_{ij} = \boldsymbol{x}_i - \boldsymbol{x}_j \qquad (2)$$

where $\boldsymbol{x}_i$, $\boldsymbol{x}_j$ and $m_i$, $m_j$ are the position and the mass of particle *i, j*, *G* is the gravitational constant, and *S* is a softening parameter used to suppress any divergence of the force at $\boldsymbol{r}_{ij} \to 0$. On the other hand, the Coulomb force between charged particles is expressed by

$$\boldsymbol{F}_i = \frac{-q_i}{4\pi\varepsilon_0} \sum_j q_j \frac{\boldsymbol{r}_{ij}}{(r_{ij}^2 + S^2)^{3/2}} \qquad (3)$$

where $q_i$ and $q_j$ are the respective charges of particle *i* and *j*, and $\varepsilon_0$ is the electric constant. This equation shows that we can accelerate computations using G6 for the system including Coulomb interaction, by the replacement of $G$, $m_i$ and $m_j$ in expression (1) with $-1/4\pi\varepsilon_0$, $q_i$ and $q_j$. In the case of the simulations presented the following sections, the host computer performed integration of the equation of motion, calculation of charging process and external forces. G6 performs calculations of interaction between dust particles. This inter-particle force is calculated as Coulomb force using expression (3).

## 3. Simulation results

### 3.1. Behaviour of dust clouds in microgravity

As an example of dusty plasma simulations using G6, we have calculated movement of dust particles which are injected into a plasma under microgravity conditions. In this simulation, we have assumed capacitively coupled radio-frequency (RF) discharge as the background plasma, and micron-order radius spheres as the dust particles. Considering the size of the dust particles and the plasma conditions, we use three external forces as the dominant external forces acting on the dust particles: ion drag force, neutral gas drag force, and electrostatic confining force [7]. The magnitudes of these external forces and others, which may play a role under similar conditions, have been investigated on an experimental basis [8]. The interaction between dust particles is calculated using the Coulomb interaction shown in expression (3). The background plasma affects the space charge quasi-neutral condition, the number of charges on the dust particle, and magnitude of the external forces. In this example, we use an argon plasma with a neutral gas pressure of 0.1 torr and a plasma density of $10^{15}$ ~ $10^{16}$ m$^{-3}$. The electron and ion temperature was 1 eV and 0.05 eV, respectively. As an initial condition, dust particles were uniformly distributed within the system using random numbers, and the velocity distribution was set to the Maxwellian velocity distribution with a temperature of 300 K. Figure 2 shows the temporal evolution of the spatial distribution of the dust particles. It is seen that dust particles are pushed out of the central region, and are arranged at the outer edge of the system. This results in the appearance of a centimeter-size void inside the dust cloud. It is also observed that the dust number-density is enhanced around the void, and the dust particles form a rather stable structure. This “void formation” has been observed in some laboratory as well as microgravity experiments [5,9]. In microgravity, sufficiently large dust particles can exist within the main plasma, different from laboratory experiments, and void formation occurs easily in dusty plasmas. The void characteristics obtained in the simulations, such as the void diameter, are similar to those seen in the microgravity experiments.

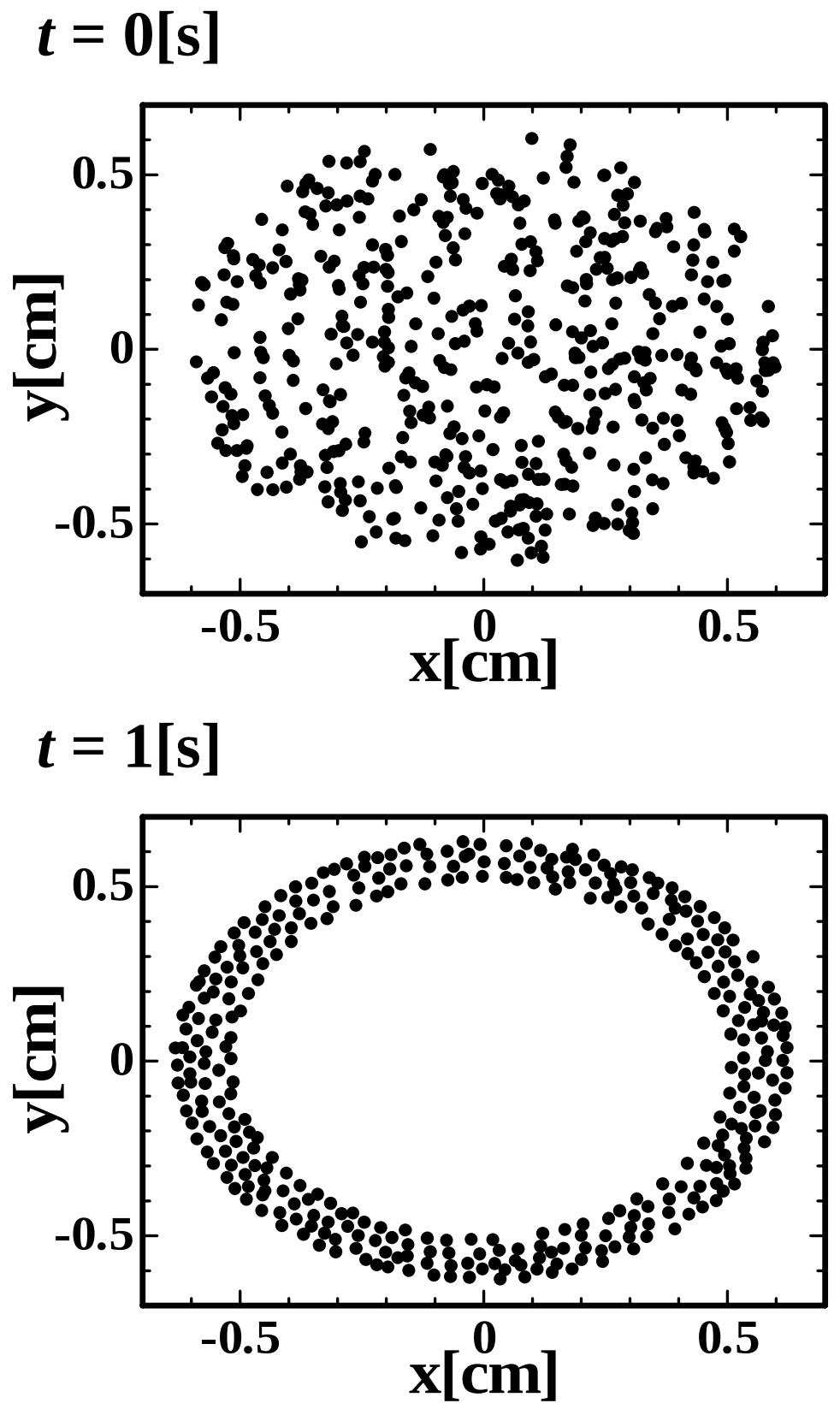

Fig.2. Cross sectional view of temporal evolutions of the dust cloud at $z$ = 0.

In the present simulation, we have also studied variations in the void diameter caused by varying external conditions. Figure 3 illustrates the variation of the void diameter as a function of the plasma density. It is seen that the void diameter increases with plasma density. However, in the region where the plasma density is very low, no void formation is observed in the simulation. This tendency is analogous with that seen in some experimental results [10], where the void size is observed to increase with input RF power.

### 3.2. Coulomb crystallization in plasmas

The next example is to simulate experiments of Coulomb crystallization in plasmas. In this simulation, we have studied crystallization of the dust clouds, due to Coulomb interactions between dust particles. These inter-particle forces are directly calculated using G6 with expression (3). We have used a capacitively coupled RF plasma for the background plasma, and a micron-order-radius sphere for the dust particles. It is assumed that the dust particles are levitating in the plasma-sheath boundary above the electrode. The external conditions, such as plasma parameters and gas pressure, are set to typical values of the usual laboratory experiments. The effects of gravity, neutral gas drag, and electric field are considered as external forces exerted on the dust particles. In the vertical direction, a strong electric field, i.e. the sheath electric field, is introduced to support the dust particles against gravity. Besides, there is a weak confining electric field in a lateral direction to confine the dust particles. In this example, we used an argon plasma with a neutral gas pressure of 0.1 torr and a plasma density of $10^{15}$ $m^{-3}$. The electron and ion temperatures were 3 eV and 0.05 eV, respectively. As an initial condition, dust particles were uniformly distributed within the system using random numbers, and the velocity distribution is set to the Maxwellian velocity distribution for a temperature of 300 K. In the simulations, the particles move about in the system with thermal velocity at first, but after a period of time, are cooled by the neutral gas and form an ordered crystalline structure. This transition from a fluid liquid-like phase to a stationary solid-like phase, is referred to as “Coulomb crystallization”. An example of the observed crystalline structure is shown in Fig.4. Figure 4 (a) shows a side view of the whole structure, and Fig.4 (b) shows a top view. From Fig.4 (a), it can be seen that the dust cloud is divided into several layers in the z-axial direction, and the dust particles are confined within each layer. Although some disorder is found in the central region, a nearly isotropic ordered structure is seen in Fig.4 (b). In some laboratory experiments, the structure of the Coulomb crystal has been reported as containing several layers arranged vertically, with the dust particles within the layers showing a uniformly-spaced isotropic arrangement in a lateral direction [3]. Unlike the laboratory experiments, the lateral arrangement of dust particles within the layers was not completely isotropic in the simulation. However, the overall structure is analogous with that of the Coulomb crystal observed in the laboratory experiments.

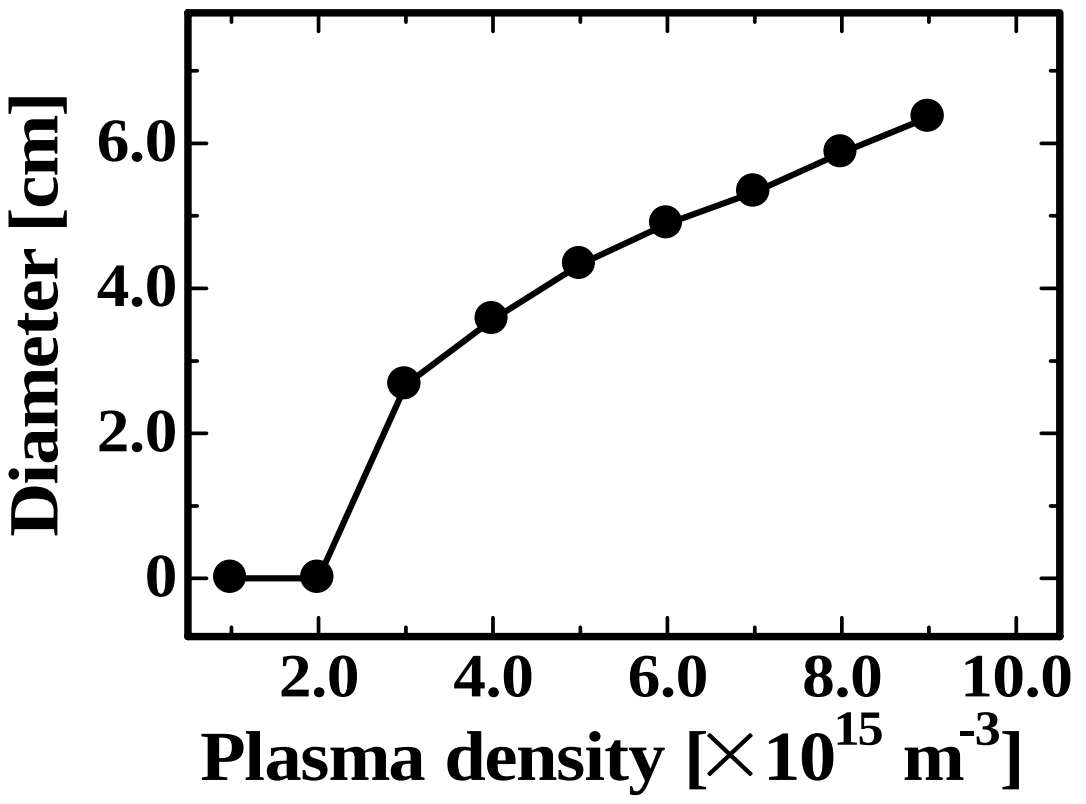

Fig.3. Dependence of the void diameter on plasma density.

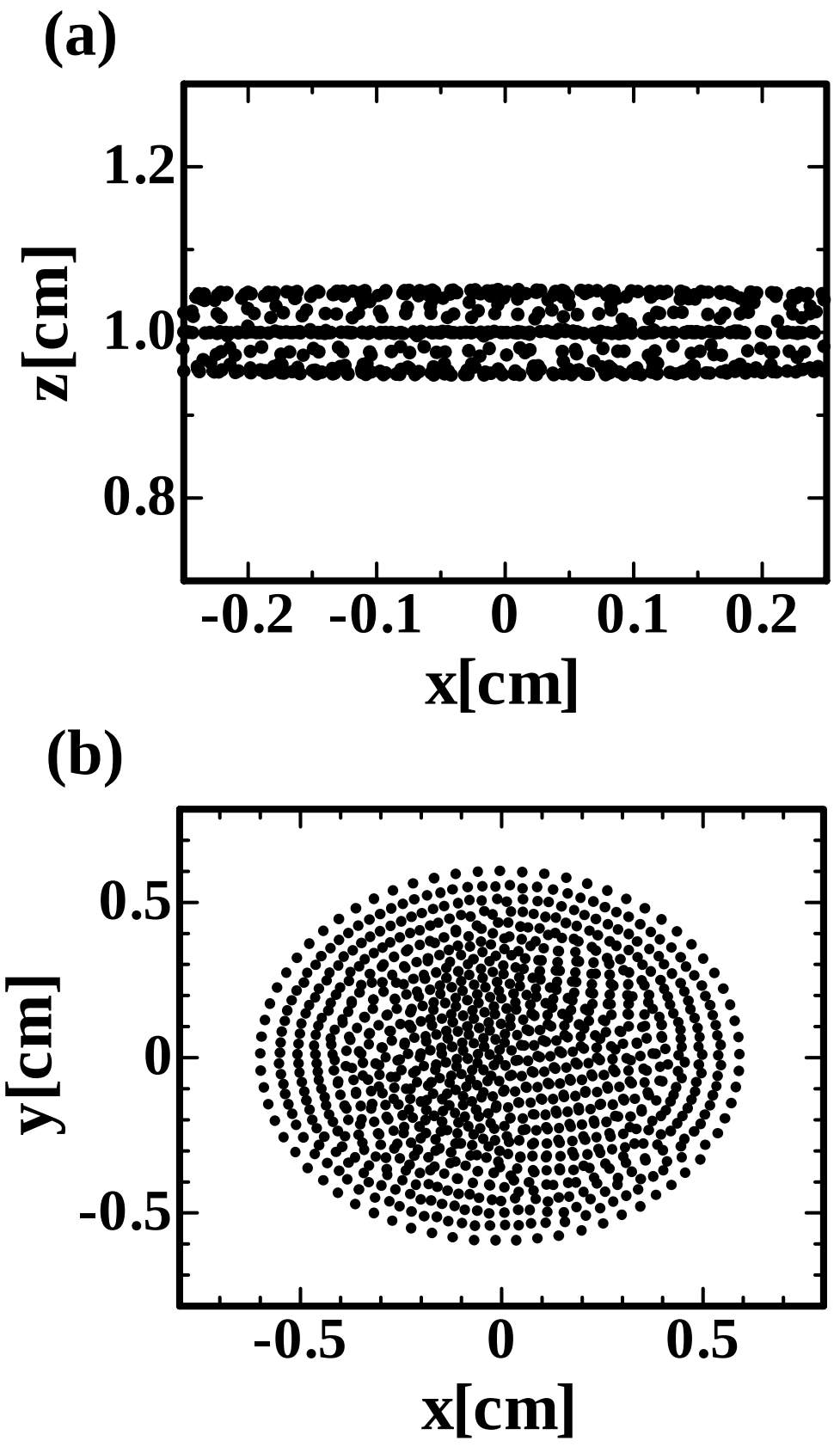

Fig.4. Crystalline structure of the dust particles. (a) Side view. (b) Top view.

### 3.3. Evaluation of the calculation speed of G6

Finally, in order to evaluate the actual calculation speed of G6 for the simulations, we compare the calculation speed of G6 and a personal computer. Figure 5 shows the computing time required to perform the whole operation of a simulation using G6, and time using only the host computer (incorporating a Pentium 4-1.6 GHz CPU). The simulation of a dusty plasma in microgravity, which is presented above, is used for the comparison. It is shown that it takes 12.1 hours to perform the simulation with $3\times10^4$ particles using the Pentium 4 personal computer, and 43 seconds using G6. This advantage becomes obvious in the case of simulations containing more than $10^4$ particles. It was also shown that the whole calculation of the simulation can be performed at an average speed of 800 Gflops by using the special- purpose computer G6.

## 4. Conclusions

We have performed simulations of dusty plasmas using a special-purpose computer system, the GRAPE-6. The collective behaviour of dusty plasmas in microgravity was simulated. It was observed that a centimeter-size void appeared inside the dust cloud under typical plasma conditions in some micro- gravity experiments. The dust particles around the void were strongly coupled and formed a stable structure. The simulated void diameter increased with plasma density. However, when the plasma density was very low, no void formation was observed. This behaviour of the dust cloud is similar to those shown by some experimental results. We have also simulated the Coulomb crystallization in plasmas. The crystalline structure of the dust particles is observed in the plasma-sheath boundary. It was seen that the crystalline structure consisted of several layers arranged vertically, with a nearly isotropic lateral arrangement of the dust particles within each layer. The crystalline structure thus obtained is analogous to that seen in some laboratory experiments. It was confirmed that the whole calculation of the simulation can be performed at an average speed of 800 Gflops by using the special-purpose GRAPE-6 computer.

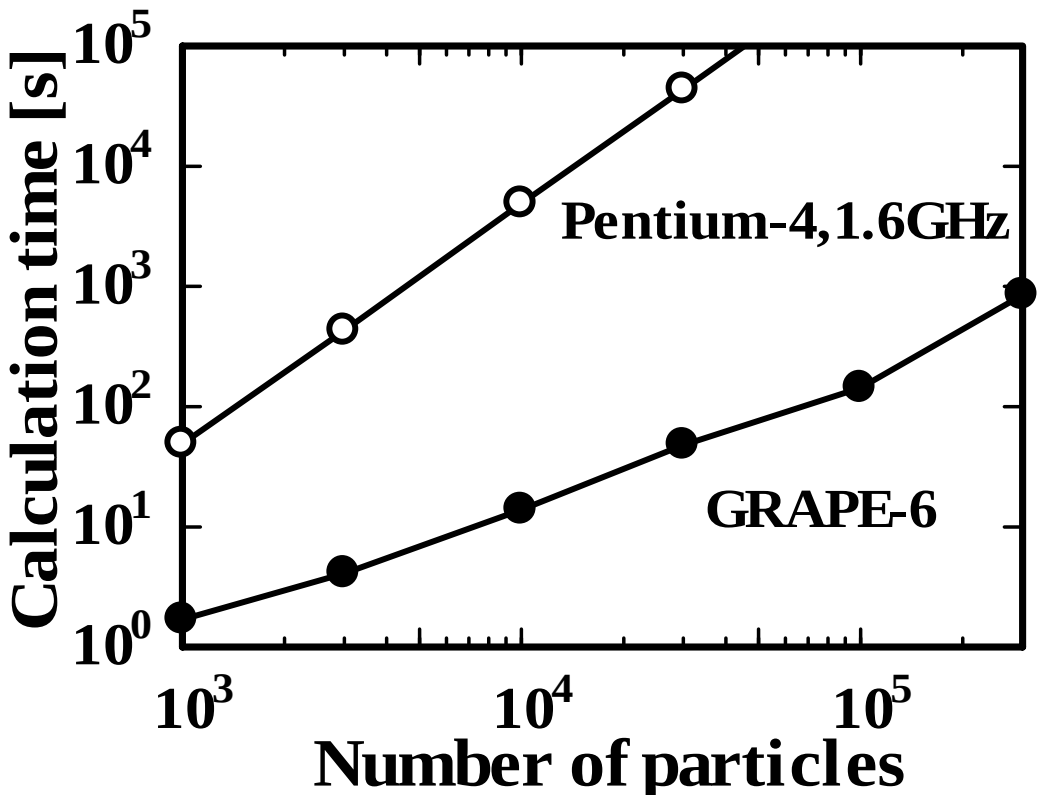

Fig.5. Comparison of the calculation time using GRAPE-6 with the time required by a Pentium-4 1.6GHz CPU.

## Acknowledgement

This work is supported in part by a grant from the Japan Small and Medium Enterprise Corporation.

## References

[1] J. Makino and Y. Funato, Astron. Soc. Japan **45,** 279 (1993).

[2] H. Thomas, G.E. Morfill, V. Demmel, J. Goree, B. Feuerbacher and D. Möhlmann, Phys. Rev. Lett. **73**, 652 (1994).

[3] Y. Hayashi and K. Tachibana, Jpn. J. Appl. Phys. **33**, L804-L806 (1994).

[4] G. Uchida, S. Iizuka, N. Sato, IEEE Trans. Plasma Sci. **29**, 274 (2001).

[5] G.E. Morfill, H.M. Thomas, U. Konopka, H. Rothermel, M. Zuzic, A. Ivlev, and J. Goree, Phys. Rev. Lett. **83**, 1598 (1999).

[6] J. Goree, G.E. Morfill, V.N. Tsytovich, and S.V. Vladimirov, Phys. Rev. E **59**, 7055 (1999).

[7] J. Perrin, *Dusty Plasmas*, edited by A. Bouchoule (John Wiley & Sons Publ., 1999) Chap. 1.

[8] C. Zafiu, A. Melzer, and Piel, Phys. Plasmas **9**, 4794 (2002).

[9] G. Praburam and J. Goree, Phys. Plasmas **3**, 1212 (1996).
[10] R.P. Dahiya, G.V. Paeva, W.W. Stoffels, E. Stoffels, G.M.W. Kroesen, K. Avinash, and A. Bhattacharjee, Phys. Rev. Lett. **89**, 125001-1 (2002).